\documentclass[useAMS,usenatbib]{mn2e}

\title[Neutron-star fall-back disks]{Interaction and ablation
 of fall-back disks in isolated neutron stars}
\author[P. B. Jones]{P. B. Jones\thanks{E-mail:
p.jones1@physics.ox.ac.uk}\\
University of Oxford, Department of Physics, Denys Wilkinson Building, 
Keble Road, Oxford OX1 3RH, England\\}
\begin{document}

\date{}

\pagerange{\pageref{}--\pageref{}} \pubyear{}

\maketitle

\label{firstpage}

\begin{abstract}
An analysis of ablation processes is made for a fall-back disk with
inner and outer radii external to the neutron-star light cylinder.
The calculated ablation rate leads, with certain other assumptions, to a
simple expression relating the inner radius and mean mass per unit
area of any long-lived fall-back disk.  Expressions for the torque
components generated by interaction with the pulsar wind are obtained.
It is not impossible that these could be
responsible for small observable variations in pulse shape and
spin-down rate but they are unlikely to be the source of the periodic
changes seen in several pulsars.

\end{abstract}


\section{Introduction}

A simple model for neutron-star disk formation from supernova fall-back
was described nearly twenty years ago (Michel 1988).  More recently, there
has been a revival of interest in the subject in connection with topics
such as planetary formation (Lin, Woosley \& Bodenheimer 1991; Wolszczan
\& Frail 1992; Miller \& Hamilton 2001), pulsar spin
evolution (Menou, Perna \& Hernquist 2001a; Blackman \&
Perna 2004 ), the anomalous X-ray pulsars (Chatterjee, Hernquist \&
Narayan 2000; Marsden et al 2001; Ek\c{s}i \& Alpar 2005; Ek\c{s}i, Hernquist
\& Narayan 2005), and a common framework for the anomalous X-ray pulsars,
soft gamma-ray repeaters and dim isolated thermal neutron stars (Alpar 2001).

Fall-back disks in isolated neutron stars may differ from the accretion disks
($\alpha$-disks) of binary
systems in a number of respects.  It is possible for them to be
at radii beyond the light cylinder radius $R_{LC}$, with
temperatures observable in the infra-red resulting from heating by X-rays
and by the pulsar wind rather than mass transfer and viscous evolution. 
A significant fraction of
the mass may be in the form of dust grains. Computations of the optical and
infra-red emission expected from fall-back disks have been made by a number
of authors (Foster \& Fischer 1996; Perna,
Hernquist \& Narayan 2000), and there have also been several experimental
searches
for the infra-red excess characteristic of dust grains
(see Bryden et al 2006).  In this way, upper limits of the
order of $10^{-2}M_{\odot}$ have
been obtained for disk masses in a number of radio pulsars
(Phillips \& Chandler 1994; L\"{o}hmer, Wolszczan \& Wielebinski 2004).  

The recent observation of an excess in the case
of the anomalous X-ray pulsar 4U 0142+61 (Wang, Chakrabarty \& Kaplan 2006)
has been interpreted by these authors as evidence for an X-ray heated
fall-back disk. The estimated mass of the disk is quite small,
$M_{d}\sim 10^{-5}M_{\odot}$, and a factor favouring its observation is
the high X-ray luminosity ($\sim 10^{36}$
erg s$^{-1}$) of the central neutron star relative to the typical
radio pulsar. Its estimated present inner radius is well outside the light cylinder,
$r_{i} \approx 4.7 R_{LC}$. An alternative analysis by Ertan et al (2007)
fits the complete infra-red and optical spectrum of 4U 0142+61
by the emission of a viscous gaseous disk with $r_{i} < R_{LC}$ and an
extinction to that object identical with the estimate made independently by
Durant \& van Kerkwijk (2006a).  This type of disk is much more complex than
the one considered in the present paper and should properly be regarded as
forming a  boundary condition for the magnetosphere (see, for example,
Cheng \& Ruderman, 1991).  The ablation
processes considered in this paper would remain relevant at $r > R_{LC}$
for this type of disk but their
use to define the inner radius $r_{i}$, as in Section 2.4, would not be
possible.

Details of the fall-back disk formation are not described here.
Initially, the disk must have been internally ionized and viscous, with
mass and angular momentum transfer rates intrinsic to the formation process.
We assume that the essential ideas of the evolutionary path described by
Menou, Perna \& Hernquist
(2001b) are valid.  The thermal ionization instability, in which
free electron recombination causes a sudden decrease in opacity, occurs
first at the outer disk radius but then propagates very rapidly inward.  We
accept the conclusion of Menou et al that a neutral, passive, gaseous disk
is then formed, with no obvious source of internal viscosity.  The power
input to the disk, considered here in Sections 2.1 and 2.4, is not adequate
to maintain temperatures above the thermal ionization instability
temperature and so does not materially affect the onset and propagation of
instability except, possibly, to delay it.

To summarize, the present paper assumes that the fall-back disks with which
it is concerned are passive, thin and have internal temperatures well below
the ionization instability temperature.  Hence, following Menou et al, there
is negligible internal ionization and the viscosity of $\alpha$-disks
is not present.  We shall be concerned only with the interaction of the
pulsar wind with the disk surface regions so that it will be possible to
neglect the complex evolutionary dynamics of neutral dust and debris disks 
(see, for example, Frank, King \& Raine 2002).
There is certainly interaction between neutral dust grains but to a good
approximation we can treat the disk as a set of annuli each having a
radius $r$ and Kepler velocity $v_{K}$.  Warping of the disk then occurs,
in general, if torque components are $r$-dependent.
It is assumed here that, in
the general case, the disk plane at formation has a finite uniform tilt
angle $\beta$, but that the disk angular
momentum is more nearly parallel rather than antiparallel with the
neutron-star spin.  We consider only disks that are external to
the light cylinder,
$r_{i} > R_{LC}$, and regard their effect on the pulsar magnetosphere as
a partial termination of the pulsar wind rather than a boundary condition
to be imposed on the solution for the magnetospheric fields.  This
limitation on $r_{i}$ means that the equivalent force on the disk derived
from the Lense-Thirring effect (see, for example, Wilkins
1972) is negligible in comparison with those found in Section 3.1. 
The Robertson-Poynting effect can also be neglected.  The reason is that
the equivalent force
on the disk depends on its Kepler velocity, $v_{K} \ll c$,
but the torque components derived in Section 3.1 from
the azimuthal part of the pulsar wind are independent of this factor.

The present paper is addressed principally to three problems. These are:
(i)		an analysis of disk ablation processes and estimates of the
ablation rate and of the infra-red luminosity;
(ii)	calculation of the alignment and precessional torques acting
between disk and neutron star;
(iii)	the question of whether or not disk alignment,
counter-alignment or precession could give rise to time-variable phenomena
that are observable over intervals of no more than several years.

The last of these has been prompted by the observation of periodic
changes in timing residuals and in average-pulse characteristics
(Stairs, Lyne \& Shemar 2000; Shabanova, Lyne \& Urama 2001; Haberl
et al 2006) of several pulsars or isolated neutron stars that
appear to be most simply explained by Eulerian precession
(for a recent review of this interpretation, we refer to
Link 2006). If the existence of Eulerian precession were established
unambiguously, there would be far-reaching consequences for our
present understanding of neutron-star internal structure.  For this
reason, we attempt in Section 4 to see if residues of fall-back disks
could exist in these cases with inner radii small enough to give
variability on the observed time-scales.

\section[]{Disk ionization and ablation}

\subsection{Momentum densities in the vacuum solution}

In order to study the interaction between a disk and the pulsar wind we shall
use inertial frame cartesian coordinates in which the neutron-star spin
angular velocity ${\bf \Omega}$ is parallel with the $z$-axis.  Where
convenient, this frame is also represented by spherical polar coordinates
$r,\theta,\phi$.  The magnetic
dipole moment is $\mu = B_{o}R^{3}$, where $R$ is the neutron star radius
and $B_{o}$ the equatorial surface field.  It is at an angle $\xi$ with
${\bf \Omega}$.  There has
been much computational work on the aligned neutron-star magnetosphere, but
in the general case with $\xi\neq 0$, the work of Spitkovsky (2006), based
on relativistic force-free electrodynamics, seems to be the only
published solution. Thus
our present investigation is based on the Deutsch vacuum solution for the
electromagnetic fields of a rotating neutron star in
the form given in recent papers by  Michel \& Li (1999) and
Ek\c{s}i \& Alpar (2005). We note that there appears to
be some disagreement about a small number of near-field terms in this
solution (see also Ferrari \& Trussoni 1973; Good \& Ng 1985) but these
are not of primary concern for the torque calculations made here. Michel \&
Li also observe that the static radial electric field term, which remains
significant at the light cylinder, is in principle undetermined because
the total electric charge of the star and
magnetosphere is an unknown quantity dependent on the details of formation.
From the electric displacement ${\bf D}$ and magnetic flux density ${\bf B}$
given by this solution in the inertial frame at $r > R_{LC}$
it is simple to write down the
spherical polar components of the momentum
density,
\begin{eqnarray} 
{\bf p} = \frac{1}{4\pi c} {\bf D}\times {\bf B}.
\end{eqnarray}  
For the torque calculation described in Section 3.1, we require them
averaged over the pulsar rotation period.
They are:
\begin{eqnarray}
\langle{\bf p}_{r}\rangle = \frac{\mu^{2}\Omega^{4}\sin^{2}\xi}{8\pi c^{5}r^{2}}
\left(1 + \cos^{2}\theta\right),
\end{eqnarray}
\begin{eqnarray}
\langle{\bf p}_{\theta}\rangle = 0,
\end{eqnarray}
\begin{eqnarray}
\langle{\bf p}_{\phi}\rangle  & = &  \frac{\mu^{2}\Omega
\cos^{2}\xi\sin\theta}{6\pi c^{2}r^{5}} +  \nonumber \\
       &    & \frac{\mu^{2}\Omega}{4\pi c^{2}r^{5}}
 \left(1 + \left(\frac{r\Omega}{c}\right)^{2}\right)
\sin^{2}\xi \sin\theta.
\end{eqnarray}
The first term in equation (4) is derived from
the product of two static irrotational field components and hence gives a
solenoidal contribution to $\langle{\bf p}_{\phi}\rangle$.
Taken at face value, this represents a purely azimuthal component of
momentum density whose presence in the near field of an aligned
rotating magnetic dipole appears not implausible.  Integration
to obtain the rate of outward transfer of angular momentum
across a sphere of large radius $r$ gives the vacuum torque,
\begin{eqnarray}
\langle{\bf \Gamma}_{v}\rangle = cr^{2}\int^{1}_{-1}d(\cos\theta)
\int^{2\pi}_{0} d\phi \langle{\bf r}\times{\bf p}\rangle,
\end{eqnarray}
of which the $z$-component is the spin-down torque,
\begin{eqnarray}
\langle\Gamma _{vz}\rangle  = \frac{2\mu^{2}\Omega^{3}}{3c^{3}}\sin^{2}\xi,
\end{eqnarray}
with $\langle\Gamma _{vx}\rangle = \langle\Gamma _{vy}\rangle = 0$, as
expected in the inertial frame.  Naturally, identical results are obtained
by integrating the time-averaged torque, formed directly from the Maxwell
tensor, over the same surface. (In a frame of reference corotating with
the star, the time-averaged  components of ${\bf \Gamma}_{v}$ are all finite
and ${\bf \Gamma}_{v} \propto ({\bf \Omega}\times {\bf \mu})\times{\bf \mu}$;
see Davis \& Goldstein 1970.  Thus the direction of ${\bf \Omega}$
changes in this frame.  Therefore, in general, it also changes relative to
the dipole moment ${\bf \mu}$ which may move in this frame during pulsar
evolution in an uncertain manner.)

The relation between wind momentum density and torque expressed
by equation (5) is no more than classical mechanics and remains valid, with
$\langle{\bf \Gamma}_{v}\rangle$ replaced by the true torque
$\langle{\bf \Gamma}\rangle$,
whatever the degrees of freedom contributing to ${\bf p_{\phi}}$.  Thus
$\langle{\bf p_{\phi}}\rangle$ is necessarily finite in a physical pulsar
wind, in which
the electromagnetic fields are loaded by a density of relativistic
charged particles.
However, it is unclear whether or not $\langle{\bf p_{\theta}}\rangle$
would remain zero in a real system:  a finite value would not contribute
to $\langle{\bf \Gamma}\rangle$, whose $x$ and $y$ components would
necessarily be zero in the inertial frame. 
The effect of these terms on the interaction between disk and pulsar wind
does not seem to have been
considered previously and is significant in the case of thin dust and
debris disks similar to those found in the major planets or in
pre-main-sequence stars (see, for example, Beckwith et al 1990) which
present a very small profile to ${\bf p_{r}}$.

\subsection{Composition of the pulsar wind}

A pulsar luminosity, observed outside the light cylinder, can be
broadly divided into a wind component $\mathcal{L}_{w}$ and blackbody
radiation from the neutron-star surface in the form of an
X-ray component $\mathcal{L}_{bol}$.  The wind luminosity, which
can be estimated from $\Omega$ and the observed spin-down rate
$\dot{\Omega}$, is further
divided into a Poynting flux of electromagnetic fields
and a relativistic particle component,
$\mathcal{L}_{w} = \mathcal{L}_{em} + \mathcal{L}_{p}$.  In previous
work on disk ablation by the pulsar wind, Miller \& Hamilton (2001)
assumed $\mathcal{L}_{p}$ to be the larger wind component, and of
baryonic composition.

However, this view of wind composition appears to be at variance with
theories of particle acceleration at $r < R_{LC}$ that are widely
accepted to be at least qualitatively valid.  In the
corotating frame of reference, there may exist finite electric-field gaps
in the open magnetic flux-line regions of the magnetosphere.  The
total current flow through these is limited by the Goldreich-Julian
charge density,
$\sigma_{GJ} = -{\bf \Omega}\cdot{\bf B}/2\pi c$, where
${\bf B}$ is the magnetic flux density.
Thus the total rate of baryon loss estimated from the polar-cap area
$\pi R^{3}\Omega/c$ intersected by open magnetic flux lines and from the
polar-cap Goldreich-Julian charge density can be at most of the order of
\begin{eqnarray}
\mathcal{R}_{b} =  \frac{\mu\Omega^{2}}{ec},
\end{eqnarray}
where $\mu$ is the neutron-star magnetic dipole moment.
But the size of the finite-field gaps is limited by intense
electron-positron pair formation with the consequence that the
Goldreich-Julian current density $\sigma_{GJ}c$ must contain both
ion and positron components. In polar-cap models, particles
are typically accelerated to energies of the order of $10^{3}$ GeV
per unit charge, though greater energies may be reached in outer-gap
models.  It is also widely accepted that, as a consequence
of the gap-limitation process, large numbers of electron-positron
pairs are formed in open magnetic flux-line regions external to
the gaps. Thus the total rate of pair formation can be given as
$\mathcal{R}_{ep} = \kappa\mathcal{R}_{b}$, but even its
order of magnitude, $\kappa \sim 10^{3-5}$, is not well known.
The energies of these pairs are several orders of magnitude lower than
the typical gap energy of $10^{3}$ GeV.  It is obvious that the
particle luminosity, estimated in the region of the light cylinder
from equation (7) and from the gap energy,
must be $\mathcal{L}_{p} \ll \mathcal{L}_{w}$ in most cases. 
For example, from the ATNF Pulsar Catalogue parameters
given for PSR 1828-11 (Manchester et al 2005), we find 
$\mathcal{R}_{b} = 8 \times 10^{31}$ s$^{-1}$ giving an estimated
$\mathcal{L}_{p} \approx 1.3\times 10^{32}$ erg s$^{-1}$, to be compared with
$\mathcal{L}_{w} = 3.6 \times 10^{34}$ erg s$^{-1}$. This is consistent
with the usual assumption (see, for example, Melatos \& Melrose 1996)
that the Poynting flux luminosity $\mathcal{L}_{em}$ is the larger
component of the wind at the light cylinder.  We shall assume that
this is maintained at disk radii.

\subsection{Ablation processes}

In this paper, we assume that the presence of a partially ionized disk
at $r > R_{LC}$, well outside any possible Alfv\'{e}n radius, has little
effect on fields and particle fluxes at
$r < R_{LC}$ and can be seen as a partial wind termination rather
than a modified boundary condition. The sequence of processes in disk
ablation is described in this Section.  We show first that the thermal
X-ray flux $\mathcal{L}_{bol}$ ionizes the outer regions of the disk surface
and so makes possible the conversion of the Poynting flux
$\mathcal{L}_{em}$ to ion or proton kinetic energy.  Ablation is then
the result of several possible secondary processes, including neutron
production in the interaction of the accelerated ions and protons with
disk nuclei. 

It is necessary to
find the conditions under which a disk consisting of
neutral hydrogen, atoms of mass $m_{A,Z}$, and dust grains
of radius $a$ has significant ionization. We shall assume that the disk
has small thickness $h$ and that it is internally neutral, with a
temperature well below the ionization instability temperatures
found by Menou, Perna \& Hernquist (2001b) for metal-rich compositions.
But the surface of a thin disk must have a finite temperature owing
to its interaction with the particle component of the wind momentum
density ${\bf p}_{\phi}$. As a result, a diffuse low-density surface
region is present which is exposed to the X-ray component
$\mathcal{L}_{bol}$, principally photons of average energy
$\mathcal{E} \approx 2k_{B}T_{s}$ determined by the neutron-star
surface temperature $T_{s}$.  The gravitational restoring force
on an atom, acting toward the plane of a thin disk, in terms of the
Kepler angular frequency $\Omega_{K}$, is
$m_{A,Z}\Omega_{K}^{2}w$ at height $w$ above the plane of the disk. 
Thus the depth of the diffuse surface must be of the order of
\begin{eqnarray}
\left(\overline{w^{2}_{A,Z}}\right)^{1/2} = 
\sqrt\frac{k_{B}T}{m_{A,Z}\Omega_{K}^{2}},
\end{eqnarray}
typically $\sim 10^{7}$ cm for hydrogen at $T = 10^{3}$ K and for
$\Omega_{K} = 10^{-2}$ rad s$^{-1}$.
The mean life of a neutral hydrogen
atom in this region against photoelectric ionization is
\begin{eqnarray}
\tau_ {pe} = \frac{\lambda}{m_{H}}\frac{4\pi r^{2}\mathcal{E}}
{\mathcal{L}_{bol}},
\end{eqnarray}
where $\lambda$ is the mass attenuation length for low-energy photons in
hydrogen and $m_{H}$ is the hydrogen atom mass. For a typical radio
pulsar, $\mathcal{E} = 10^{2}$ eV equivalent to
$\mathcal{L}_{bol} = 8\times10^{31}$ erg s$^{-1}$ for a neutron star
radius $R = 10^{6}$ cm.
The Particle Data Group (Yao et al 2006) give
$\lambda = 1.0 \times 10^{-4}$ g cm$^{-2}$ so that even at a radius
$r = 10^{11}$ cm, the mean life is very short, $\tau _{pe} =15$ s.
For higher
atomic numbers $Z$, lifetimes against ionization are of the same order.

The extent of dust grain sublimation is uncertain.
Dust grains  exposed to $\mathcal{L}_{bol}$ have an estimated
equilibrium temperature $T_{bol} = T_{s}\sqrt{R/2r} \approx 1300$ K, provided
details of grain composition, emissivity and shape are neglected. Sublimation
rates may be significant at this temperature which is also approximately
equal to the K-band Wien's displacement law temperature.  But its
dependence on $T_{s}$ and $r$ make it impossible to give any general
statement.  This estimate
assumes a grain radius $a \geq 10^{-4}$ cm such that almost all incident
photons interact. The mass
stopping power for low energy electrons has been tabulated by Seltzer
\& Berger (1982). It is so large that there is no doubt that
the energy of the emitted photo-electrons is contained within dust grains
of these radii.
Smaller grains, with incomplete retention of photo-electron energy, have
lower radiative equilibrium temperatures.
On the basis of the Dulong and Petit law, the thermal equilibrium
temperature is reached within times short compared with the Kepler period
of the disk. The ionization time $\tau _{pe}$ is also smaller than
the Kepler period leading to the conclusion that,
for interesting intervals of $\mathcal{L}_{bol}$ and $r$, matter in the
very diffuse surface regions of a disk has a high degree of ionization
independent of the state of the dust component.
Neutron star blackbody photons therefore have an important effect on the
state of the disk but their energies, $\mathcal{E} \sim 10^{2}$ eV,
are several orders of magnitude too small to eject protons
(a proton with $10^{2}$ eV/c momentum has negligible kinetic energy;
see also Miller \& Hamilton 2001).
  
Following the discussion of the particle component of $\mathcal{L}_{w}$
in Section 2.2, we shall assume that the fields at the disk radius,
though not necessarily at radii that are many orders of magnitude greater,
satisfy the ideal magnetohydrodynamic condition
$c{\bf E} = - {\bf v}_{0} \times {\bf B}$, where ${\bf v}_{0}$ is
the particle velocity (see, for example, Melatos \& Melrose 1996;
for a review of pulsar wind nebulae, see Gaensler \& Slane 2006).
To the extent that this is true, the further conditions $E < B$
and ${\bf E}\cdot {\bf B} = 0$ are also satisfied.  In this case,
the qualitative details of particle acceleration as a result of
the interaction of the Poynting flux in the wind
with the diffuse ionized disk surface can be
analyzed easily by making a Lorentz transformation of velocity
\begin{eqnarray}
\frac{{\bf v}^{\prime}}{c} = \frac{{\bf E}\times{\bf B}}{{\bf B}^{2}}
\end{eqnarray}
from the inertial frame of Section 2.1 to a frame with fields
${\bf E}^{\prime}, {\bf B}^{\prime}$ in which ${\bf E}^{\prime} = 0$
(see, for example, Landau \& Lifshitz 1962).  In this frame,
${\bf B}^{\prime}$ is parallel with ${\bf B}$ and of magnitude
$B^{\prime} = \sqrt{B^{2} - E^{2}}$.  A proton formed by
photoelectric dissociation of neutral hydrogen has an initial
velocity
(neglecting Kepler and thermal velocities) in this frame of
$-{\bf v}^{\prime}$ and its subsequent orbit is a
circle of radius $v^{\prime}/\omega _{B}$ in a plane perpendicular
to ${\bf B}^{\prime}$, with angular frequency
$\omega _{B} = ev^{\prime}B^{\prime}/cq^{\prime}$ where
$q^{\prime}$ is its momentum
in this frame. Thus its time-averaged velocity in the original
inertial frame is simply ${\bf v}^{\prime}$.  Although
${\bf v}^{\prime}$ and ${\bf v}_{0}$ are not, in general,
exactly parallel, it is broadly correct to say that the proton
is transported
forward, with the existing particle component of $\mathcal{L}_{w}$,
into the body of the disk. As this process continues, the Poynting
flux is  necessarily converted to kinetic energy.  The energy
transfer is linearly dependent on rest mass and thus is almost
entirely to protons or partially ionized atoms.  The penetration
depth is very roughly defined by equating
$\mathcal{L}_{em}/4\pi r^{2}c$
with the particle pressure, as for the collisionless shock
present in pulsar wind nebulae (Rees \& Gunn 1974).

In this elementary analysis, the time-dependence of the
${\bf E}$ and {\bf B} fields in the inertial frame
has been ignored.  This appears
a reasonable approximation because, at disk radii $r \sim R_{LC}$,
the proton $\omega_{B}$ is several orders of magnitude greater than
$\Omega$.  We emphasize that the diffuseness of the disk surface is
the important factor in its interaction with the Poynting flux
component of the pulsar wind.

Disk ablation, which we define as baryon loss, can occur principally
through three processes. The initial assumption made here is that the
most important is the production of
neutrons in strong interactions between the accelerated protons
or with heavier nuclei. The nuclear interaction cross-section for
baryons is defined primarily by the nuclear radius.  Thus the
mass attenuation length is $\lambda_{b} \approx 100$ g cm$^{-2}$.
A second source of loss is the possibility that, with the reflection
of ${\bf E}$ and ${\bf B}$ fields that is neglected here,
protons and ions may be transported outward and away from the surface.
Finally, the secondary protons formed in relativistic proton
interactions inside the disk can also acquire  
a momentum component parallel or antiparallel with ${\bf B}$
which allows them to move out of the disk in a time small
compared with the neutron star rotation period. These processes are
all independent of the state of the dust component in the disk.The order of
magnitude ablation rates obtained from the first of them
are very different from those
of Miller \& Hamilton (2001) which, apart from a solid angle
factor, were simply equal to
$\mathcal{L}_{w}$ divided by the gravitional potential energy of a
disk proton.

Neutron production in strong interactions in the body of the disk
has characteristics similar to those found by
Agosteo et al (2005) who have measured the neutron spectra given by
the interaction of 40 GeV/c protons with a copper target.  (That the
nuclear charge $Z=29$ exceeds the disk average is unimportant for
this purpose.)
There are two groups of neutrons.  Qualitatively, forward-directed
neutrons interact with disk nuclei,
as do secondary protons and mesons, to produce a hadronic cascade
with length scale determined by $\lambda_{b}$.
The more numerous neutrons are from decay of target nuclei
and are emitted isotropically with average energy $\sim 3$ MeV.
These neutrons interacting with disk nuclei are either moderated by
elastic scattering or undergo $(n,\gamma)$ capture.  The
photon component of the cascades, mostly from $\pi^{0}$
meson decay, can be a source of neutrons through excitation of
the giant dipole state in nuclei, but this is unimportant compared
with direct strong-interaction production.  Provided the disk is
optically thick, the photon energy passes to thermal degrees of
freedom through Compton scattering and photoelectric absorption.
Very qualitatively, we can assume that, in a high-energy
cascade, protons of momentum less than
1 GeV/c do not interact further but come to rest after losing
energy by ionization in accordance with the Bethe-Bloch formula
(see Yao et al 2006).  For these considerations, it does not matter
whether the target nuclei in the body of the disk, where the
${\bf E}$ and ${\bf B}$ fields are much reduced by conversion of the
Poynting flux to kinetic energy, are free or still bound as atoms
in dust grains.

For disks that are optically thick for high energy protons,
satisfying $\rho h \gg \lambda _{b}$, where $\rho$ is the
matter density, diffusion of the low-energy group of neutrons
backward to the surface is the important mechanism of ablation.
Obviously, this process is dependent on the angle $\chi$ between
${\bf p}_{\phi}$ and the disk outward normal unit vector
${\bf n}_{\perp}$. 
The neutron production rate grows approximately exponentially with
depth, the scale length being $\sim \lambda_{b}\sin\chi$, but the
probability of diffusion to the surface without capture
decreases more rapidly.  It appears that only the primary and
a small number of secondary target nuclei in a cascade are effective
sources of neutrons.  On this basis, and the neutron production
measurements of Agosteo et al, we shall adopt a number
$\mathcal{N}$, in the interval $1<\mathcal{N}<10$, as the mean
number of neutrons escaping the disk per proton accelerated
by the incident Poynting flux as described above.  We shall
also assume a constant value $\mathcal{E}_{p} = 10$ GeV for the
accelerated energy.

Neutron diffusion is more probable from disks that are not
optically thick to high-energy protons, for example, those with
$\rho h \sim 10 \lambda_{b}$.  Thus the ablation rates of
these can be higher, as noted by Miller \& Hamilton (2001).
There is also the interesting possibility in these
instances that the net momentum transfer normal to the disk may be
antiparallel with to the normal component of
$\langle{\bf p}_{\phi}\rangle$.  The arguments for this will be
given in Section 3.3 with a brief discussion of some consequences
that might be observable.

It must be obvious that the analysis of ablation processes
given here is intended to be no more than qualitative in the
extreme.  In particular, there is no rigorous solution for the
interaction of the ${\bf E}$ and ${\bf B}$ fields with the
diffuse outer regions of the disk.  One difficulty is that the
value of $B$ at the light cylinder varies by four orders of
magnitude for the specific neutron stars considered here. Thus the
description of Poynting flux conversion to particle kinetic energy
given here is more obviously valid for the smaller values of the
orbit radii $v^{\prime}/\omega_{B}$ than for the larger values
present in long-period neutron stars such as 4U 0142+61.
But these failings can be
viewed less seriously when it is remembered that 
the properties of the physical pulsar wind, even in the absence
of a disk, are not well known.  The present analysis attempts
no more than to show that, on interaction with the disk, much of
the Poynting flux is irreversibly converted to proton kinetic
energy and that the average proton momentum is roughly parallel
with the Poynting vector. 
  
\subsection{The ablation rate and disk luminosity}

We consider first the effect of the time-averaged azimuthal
pulsar wind component $\langle{\bf p}_{\phi}\rangle$ on the disk.
The coordinate system adopted is that defined in Section 2.1.
and the disk is assumed initially plane, with uniform tilt
angle $\beta$ and rectilinear line of nodes in the $xy$-plane
at an angle $\gamma$ with the $x$-axis.  For integration over
the area of the disk it is convenient to transform from
spherical polar coordinates $\theta,\phi$ to an
azimuthal angle $\psi$, defined in the disk plane and with
respect to the line of nodes. The angular relations satisfied
by points on the disk are:
\begin{eqnarray}
\cos\chi    &  =  & - \cos(\phi - \gamma)\sin\beta, \nonumber \\
\cos\theta  &  =  & \sin\beta\sin\psi, \nonumber \\
\sin(\phi-\gamma)  &  =  & \frac{\cos\beta\sin\psi}
{\sqrt{1 - \sin^{2}\beta\sin^{2}\psi}},  \nonumber \\
\cos(\phi-\gamma)   &  =  &  \frac{\cos\psi}
{\sqrt{1 - \sin^{2}\beta\sin^{2}\psi}}.
\end{eqnarray}
The power input per unit area of disk is
$-c^{2}\langle p_{\phi}\cdot{\bf n}_{\perp}\rangle$.  
For a disk annulus of radius $r$ and width $\delta r$, the
total power input to both sides is,
\begin{eqnarray}
\delta \mathcal{W} = r\delta r\int^{2\pi}_{0} d\psi
c^{2}\left|\langle {\bf p}_{\phi}\cdot
 {\bf n}_{\perp}\rangle\right|.
\end{eqnarray}
It will be convenient to eliminate $\mu$ in favour of
$\mathcal{L}_{w} = \Omega \langle\Gamma_{vz}\rangle$ using
equations (2), (4) and (6).  Given that
\begin{eqnarray}
\int^{2\pi}_{0}d\psi\sin\theta\left|\cos(\phi - \gamma)\right| = 4,
\end{eqnarray}
equation (12) can be re-expressed as,
\begin{eqnarray}
\delta\mathcal{W} & = &  r\delta r\frac{3c^{3}\mathcal{L}_{w}
\sin\beta}{2\pi\Omega^{3}\sin^{2}\xi}  \nonumber  \\
   &   & \left(\frac{2}{3r^{5}}\cos^{2}\xi + \left(\frac{1}{r^{5}} + 
\frac{\Omega^{2}}{r^{3}c^{2}}\right)\sin^{2}\xi\right). 
\end{eqnarray}
The first term on the right-hand side of this equation derives
from the solenoidal component of $\langle{\bf p}_{\phi}\rangle$
which was mentioned in Section 2.1.  But its singularity at
$\xi = 0$ is clearly an artefact of the unphysical vacuum
solution for the ${\bf E}$ and ${\bf B}$ fields and
for disks external to the light cylinder with radii
$r_{i} < r < r_{o}$, we need retain only the final term in
equation (14).  Integration gives the total estimated power input
to the disk from the azimuthal wind component,
\begin{eqnarray}
\mathcal{W} = \frac{3}{2\pi}R_{LC}\mathcal{L}_{w}\sin\beta
\left(\frac{1}{r_{i}} - \frac{1}{r_{o}}\right).
\end{eqnarray}
The rate of baryon loss per unit area of disk at radius $r$ is,
\begin{eqnarray}
\mathcal{B} = \frac{3R_{LC}\mathcal{N}\mathcal{L}_{w}\sin\beta}
{4\pi^{2} r^{3}\mathcal{E}_{p}}.
\end{eqnarray}
Evaluation of equations (15) and (16) for the estimated disk
parameters of 4U 0142+61 (Wang et al 2006) is of interest.
The wind luminosity derived from the ATNF parameters
(Manchester et al 2005) is $1.2 \times 10^{32}$ erg s$^{-1}$
and the inner and outer radii are $4.7R_{LC}$ and $16R_{LC}$,
respectively. The total power input to the disk is then
$\mathcal{W} = 9 \times 10^{30} \sin\beta$ erg s$^{-1}$. At the
inner radius, the power input per unit area is
$5\times 10^{7}\sin\beta$ erg cm$^{-2}$ s$^{-1}$ which is
large in relation to the gravitational potential energy per unit
area of disk divided by the $t_{0}\sim 10^{5}$ yr lifetime of the
neutron star, $GM\bar{\Sigma}/t_{0}r_{i} \sim 5\times 10^{6}$
erg cm$^{-2}$ s$^{-1}$,
where $M$ is the neutron star mass and
$\bar{\Sigma} \approx 1.6 \times 10^{4}$ g cm$^{-2}$ is the mean
mass per unit area
found from the disk mass and radii given by Wang et al.
Thus the inner edge power input from the wind is large compared
with any
possible rate of thermal energy release by viscous evolution
of the disk. Assuming
re-radiation exclusively in the infra-red, this power input
corresponds with a blackbody temperature of
$T \le 960$ K, which is lower than the
temperature of $1200$ K
found by Wang et al.  This is not unreasonable owing to the
very large value of $\mathcal{L}_{bol} \gg \mathcal{L}_{w}$
for this neutron star.  We have assumed the disk thickness to be
small and uniform.  But it is possible that it has some
radial dependence, possibly of the form $h\propto r^{9/7}$ which
is characteristic of
point-source illumination of an
$\alpha$-disk (see, for example, Pringle 1996), and that some
of the power input to the disk is from $\mathcal{L}_{bol}$
as in the model fit of Wang et al. The model adopted by these
authors must include some assumption about the absolute
thickness of the disk, in effect, the solid angle it subtends
at the neutron star.  This does not appear to be stated, but it
is also of interest that their fit involves a very large X-ray
albedo, $\eta_{d} = 0.97$. Finally, we have to accept that
non-disk contributions to the the infra-red luminosity may also be
significant.  It is possible that
the recent observation of short time-scale downward fluctuations
in the K-band intensity of this object (Durant
\& van Kerkwijk 2006b), not readily accommodated within the passive
disk model, may be evidence for these.

Evaluation of equation (16) at the present value of $\Omega$
gives an estimated baryon loss rate, at the
inner radius, of
$3\times 10^{10}$ cm$^{-2}$ s$^{-1}$ for $\mathcal{N} =10$ and
$\mathcal{E}_{p} = 10$ GeV, which is negligible in comparison
with the mean disk density $\bar{\Sigma}$.  This is an essential
result because the ablation rate must have been considerably
higher at earlier times following neutron-star formation.
It is more interesting to integrate the ablation rate over the
life of the star, assuming simple spin-down given
by equation (6), and to compare the total baryon loss per
unit area at the inner disk radius $r_{i}$ with $\bar{\Sigma}$.
We have,
\begin{eqnarray}
\int^{t}_{0}\mathcal{B} dt = \frac{3cI}{4\pi^{2}r^{3}}
\left(\frac{\mathcal{N }\sin\beta}{\mathcal{E}_{p}}\right)
\left(\Omega_{0} - \Omega\right),
\end{eqnarray}
where $I$ is the neutron star moment of inertia and
$\Omega_{0}$ its spin angular frequency at formation.
This assumes that, at earlier times, the disk extended inward to
smaller radii but has since suffered ablation to its present
value of $r_{i}$.  It
provides an estimate of a combination of the unknown
parameters $\mathcal{E}_{p}$ and $\mathcal{N}$ in terms of
quantities accessible to experimental measurement,
\begin{eqnarray}
\frac{\mathcal{N}\Omega_{0}\sin\beta}{\mathcal{E}_{p}}
 = \frac{4\pi^{2}r^{3}_{i}\bar{\Sigma}}{3cIm_{H}}. 
\end{eqnarray}
The comparison for 4U 0142+61 gives a value for the right-hand
side of equation (18) of $3.4\times 10^{7}$ erg$^{-1}$ s$^{-1}$
or $5\times 10^{4}$ rad s$^{-1}$ GeV$^{-1}$. 
Insertion of the values we adopted in Section 2.3 for
$\mathcal{N}$ and $\mathcal{E}_{p}$
then leads to $\Omega_{0}\sin\beta = 5\times 10^{4}$ which is
obviously about two orders of magnitude too large.  But given
the considerable uncertainties both in our parameter values
and in those estimated by Wang et al, the discrepancy is not
disturbing.  If the disk parameters of Wang et al are accepted,
the comparison suggests smaller $\mathcal{E}_{p}$ and larger
$\mathcal{N}$ than the values adopted in Section 2.3.
Possibly, the two processes of baryon loss listed but not considered
there may be more important than we assumed.  We have also neglected
any increase or decrease in disk radii caused by angular momentum
transfer from the neutron star via the torque component
${\bf \Gamma}_{\perp}$ given by equation (24) in Section 3.1.

Apart from the uncertainties in $\mathcal{N}$ and
$\mathcal{E}_{p}$, we emphasize that there are at
least two principal sources of uncertainty in our
calculation of $\mathcal{W}$.  Our treatment of the wind-disk
interaction in Section 2.3 is elementary and does not give
any estimate of the reflection of wind energy.  There is also
the uncertainty in our assumption that ${\bf p}_{\theta}$ makes
no contribution because its time-average vanishes. This
may not be too unsatisfactory for torque estimates, as in
Section 3.1, but may nevertheless introduce an error in
calculating the total power input.

\section[]{Disk-pulsar torques}

\subsection{Calculation of the torque}

Calculation of the torque components acting on the disk
follows that for $\delta\mathcal{W}$ given by equations (11)
and (12).  It will be convenient consider an annulus of width
$\delta r$ and to resolve the torque acting on it in
the plane of the disk into components $\delta{\bf \Gamma}_{p}$
and $\delta{\bf \Gamma}_{a}$ that are, respectively, parallel
with and perpendicular to the line of nodes.  The components
are,
\begin{eqnarray} 
\delta\Gamma_{p} =  - cr^{2}\delta r\int^{2\pi}_{0}d\psi
\sin\psi\langle p_{\phi}\rangle\left|\cos\chi\right| \nonumber \\
\left(\cos\chi + \frac{2}{3}{\rm sgn}(\cos\chi)\right),
\end{eqnarray}
and
\begin{eqnarray}
\delta\Gamma_{a} = cr^{2}\delta r\int^{2\pi}_{0}d\psi
\cos\psi\langle p_{\phi}\rangle\left|cos\chi\right| \nonumber \\
\left(\cos\chi + \frac{2}{3}{\rm  sgn}(\cos\chi)\right),
\end{eqnarray}
formed from the momentum flux perpendicular to the plane
(neglecting any wind-energy albedo) and
with the assumption that the disk is optically thick and
re-radiates photons as a Lambertian surface.  Use of the
time-averaged azimuthal momentum density in these expressions
is valid because the neutron star rotation period is several
orders of magnitude smaller than the Kepler period of
the disk.  Thus it is permissible to treat the annulus, though
not the whole disk, as a rigid body.
Evaluation of the first of these integrals gives a
zero disk precession torque,
$\delta\Gamma_{p} = 0$.  This cancellation of contributions from
the four quadrants of $\psi$ is a consequence of the vacuum-field
form of $\langle p_{\phi}\rangle$, as given by equation (4), and
it is by no means obvious that it would hold for a physical
pulsar wind.  Integration of equation (20) gives a finite
torque which acts to align the angular momentum
$\delta{\bf L}_{d}$ of the disk annulus with the neutron-star spin,
\begin{eqnarray}
\delta\Gamma_{a} = cr^{2}\delta r\left(\frac{3c\mathcal{L}_{w}}
{8\pi\Omega^{3}\sin^{2}\xi}\right)\left(4I_{a}\sin^{2}\beta
+ \frac{2}{3}\pi\sin\beta\right) \nonumber \\
\left(\frac{2}{3r^{5}}\cos^{2}\xi + \left(\frac{1}{r^{5}} + 
\frac{\Omega^{2}}{c^{2}r^{3}}\right)\sin^{2}\xi\right),
\end{eqnarray}
in which the integral,
\begin{eqnarray}
I_{a} = \int^{\pi/2}_{0}d\psi \frac{\cos^{3}\psi}
{\sqrt{1 - \sin^{2}\beta\sin^{2}\psi}},
\end{eqnarray}
is a slowly varying function of $\beta$,
$0.66 < I_{a} < 0.79$.  

The torque component perpendicular to the plane of the disk is,
\begin{eqnarray}
\delta\Gamma_{\perp} = cr^{2}\delta r\int^{2\pi}_{0}d\psi
\langle p_{\phi}\rangle\left|\cos\chi\right| \nonumber \\
\left(\cos\beta\sin\theta + 
\sin\beta\cos\theta\sin(\phi-\gamma)\right).
\end{eqnarray}
Evaluation of the angular integral using equations (11) gives,
\begin{eqnarray}
\delta\Gamma_{\perp} =  cr^{2}\delta r\left
(\frac{3c\mathcal{L}_{w}}
{8\pi\Omega^{3}\sin^{2}\xi}\right)\left(2\beta\cos\beta
\left(1 + \sin^{2}\beta\right)\right)  \nonumber  \\
\left(\frac{2}{3r^{5}}\cos^{2}\xi + \left(\frac{1}{r^{5}} +
\frac{\Omega^{2}}{c^{2}r^{3}}\right)\sin^{2}\xi\right).
\end{eqnarray}
We shall assume that the angular momentum of a fall-back disk
is likely to be more nearly parallel, than anti-parallel, with
the neutron star spin, in which case the interaction increases
$\delta L_{d}$  and moves the disk outward to larger radii.

\subsection{The effect of torques on the neutron star}

The motivation for considering this aspect of the interaction
is that the angular momentum ${\bf L}$ of the
neutron star can be several orders of magnitude smaller than
that of the disk.  In the case of 4U 0142+61, for example, the
disk parameters estimated by Wang et al, including its very
small mass $M_{d}\sim 10^{-5}M_{\odot}$, give
$L_{d} = 2.0 \times 10^{47}$ g cm$^{2}$ s$^{-1}$ which is
between one and two orders of magnitude larger than $L$.
Consequently, the changes in direction of both ${\bf L}$
and ${\bf L}_{d}$ can be significant.  Thus the
Euler equation for the neutron-star angular momentum in the
absence of a disk, $\dot{\bf L} = {\bf \Gamma}$, is replaced by,
\begin{eqnarray}
\dot{\bf L} + \dot{\bf L}_{d} = \left(\tilde{\bf \Gamma}
- {\bf \Gamma}_{d}\right) + \left({\bf \Gamma}_{d}\right),
\end{eqnarray}
for the whole system, in which ${\bf \Gamma}_{d} =
{\bf \Gamma}_{a} + {\bf \Gamma}_{p} + {\bf \Gamma}_{\perp}$ is
the integrated torque acting on the disk,
and $\tilde{\bf \Gamma}$ is the
modified non-disk torque.  For hypothetical disks with
$r_{i} \gg R_{LC}$,
the terms in equations (21) and (24) that are significant
contributors to the torque would be
those derived from the long-range component of
${\bf p}_{\phi}$ in equation (4).  But the vacuum torque
$\langle{\bf \Gamma}_{v}\rangle$ can also be obtained from this
term, as could the true $\langle {\bf \Gamma}\rangle$ if the true
${\bf p}_{\phi}$ were known (see equation 5).  This suggests,
but does not prove, that the first, bracketed, term on the
right-hand side of equation (25)
approaches $\langle{\bf \Gamma}\rangle$ in this limit, the disk
torque being cancelled by terms in
$\langle \tilde{\bf \Gamma}\rangle$.  However, there is no
reason to suppose that this cancellation is exact at disk
radii $r\sim R_{LC}$ of physical interest where the near-field
terms in equation (4), (21) and (24) are not negligible.

The terms in ${\bf \Gamma}_{d}$, including any non-zero
${\bf \Gamma}_{p}$, are all finite in the inertial frame
when averaged over the neutron star period.
Thus the direction of ${\bf L}$ must change during the
evolution of the system, but at a rate that is unlikely to
be of observational interest.  To confirm that this is so,
we can consider PSR 1828-11 as an example and suppose that ${\bf L}$ is
at a small angle $u$ with the total angular momentum
${\bf L} + {\bf L}_{d}$ and precesses about it with angular
frequency $\Omega_{p}$.  The torque required is
${\bf \Gamma}_{d} \sim Lu\Omega_{p} = 2\times 10^{37}$
g cm$^{2}$ s$^{-2}$
for a $500$ day precession with amplitude $10^{-2}$, several
orders of magnitude greater than the spin-down torque
which is $ \mathcal{L}_{w}/\Omega = 2.3 \times 10^{33}$
g cm$^{2}$ s$^{-2}$ for this pulsar. 

\subsection{Alignment and precesssion of the disk}

The initial evolution of a very thin disk with small
$\bar{\Sigma}$ and $L_{d} \ll L$ can be obtained directly from
equation (25), to the extent that we can neglect the effect
of changes in its geometrical form on calculations of
${\bf \Gamma}_{d}$.  The solution is,
\begin{eqnarray}
\dot{L}_{d} & = &  \Gamma_{\perp} \nonumber \\
L_{d}\dot{\beta} & = & -\Gamma_{a}  \nonumber \\
L_{d}\sin\beta\dot{\gamma} & = & \Gamma_{p}.
\end{eqnarray}
From a hypothetical initial state of a plane disk with
$\beta \neq 0$, the alignment torque ${\bf \Gamma}_{a}$
reduces $\beta$ at a radius-dependent initial rate given by
\begin{eqnarray}
\dot{\beta} =  - \frac{\delta\Gamma_{a}}{\delta L_{d}}
 =  -\frac{1}{2\pi\Sigma(GM)^{1/2}r^{5/2}}
 \left(\frac{3\mathcal{L}_{w}}{8\pi\Omega}\right) \nonumber  \\
 \left(4I_{a}\sin^{2}\beta + \frac{2}{3}\pi\sin\beta\right),
\end{eqnarray}
in which only the long-range term in equation (21) has been
retained.  The disk mass per unit area $\Sigma$ in equation (27)
is the local value, not the mean $\bar{\Sigma}$.
There is initially no precession for the
vacuum-field pulsar wind assumed in Section 3.1.  But the
$r$-dependent alignment rate destroys the planar form of the
disk and therefore introduces a further torque contribution
derived from the radial momentum flux
$c\langle p_{r}\rangle + \mathcal{L}_{bol}/4\pi r^{2}c$.
This is the torque concerned in the Pringle instability
(Pringle 1996; see also Petterson 1977, Frank et al 2002).
Initially, it is a contribution to the precessional
component ${\bf \Gamma}_{p}$, but further evolution of the
shape and motion of the disk is more complex.  It is worth
adding that any finite torque component derived from equation (19)
as the consequence of a
momentum density $\langle{\bf p}_{\theta}\rangle \neq 0$
in a physical wind, rather than the vacuum field solution
used in Section 2.1, would also induce precession.

In relation to disk alignment and precession, the interesting
question is whether or not these processes are sufficiently
rapid to produce time-varying phenomena that are observable
within periods of no more than several years.  For phenomena
involving
the body of a $10^{-5}M_{\odot}$ disk, the answer must be in
the negative because $\Gamma_{d}$ is perhaps an order of
magnitude smaller than the spin-down torque $\Gamma_{vz}$
yet, as we have noted in Section 3.2, the disk angular momentum
$L_{d}$ is at least of the same order as $L$.

But the inner edge of a disk must have small local values of
$\Sigma$ owing to its continual ablation by baryon loss,
given by equation (16).  In principle, the alignment rate
given by equation (27) for these radii could become observably
fast.  There is also the likelihood, mentioned
briefly in the penultimate paragraph of Section 2.3, that
for some intervals of small $\Sigma$, the
net momentum transfer normal to the disk is antiparallel with
the normal component of $\langle{\bf p}_{\phi}\rangle$.

We can see how this arises by referring back to both the
description of hadronic cascade development in Section 2.3 and
the torque calculations of Section 3.1.  The integrands in
equations (19) and (20) both contain the rate of momentum
transfer normal to the disk and per unit area.  Both expressions
are based on two assumptions; that the disk re-radiates as a
Lambertian surface and
that it does so from the surface of incidence with which the
wind interacts.
The second assumption is certainly valid if the disk is optically
thick for the photon spectrum emitted and also has sufficient
thickness, at least of the order of
$\Sigma \sim 10^{2}\lambda_{b}|\cos\chi|$, that the
hadronic cascades terminate much nearer the surface of incidence
rather than the
far surface.  But the transfer of energy in a cascade from the
primary particle to thermal degrees of freedom and to low-energy
photons of perhaps $10^{2}-10^{4}$ eV occurs predominantly
in the late stages
of its development.  Thus for
$\Sigma \sim 10\lambda_{b}|\cos\chi|$,
the wind heats the far surface of the disk, not the surface of
incidence.  The effect of this is to change the sign
of the $sgn(\cos\chi)$ terms in equations (19) and (20) and,
unless $\chi$ is relatively small, to change the sign of
$\delta\Gamma_{a}$, transforming it
from an alignment to a counter-alignment torque.  There is
a second factor having the same effect.  The production of
low-energy neutrons referred to in Section 2.3  also occurs
mainly in the late stages of cascade development.  For these
$\Sigma$, the baryon loss rate per unit area much exceeds
that given by equation (16) and is mainly through the far surface
of the disk.  This is significant because massive low-energy
particles are effective carriers of momentum.

Validation of these qualitative arguments would require very
extensive Monte Carlo calculations of cascade development.
But although these have not been made, we suggest that there
is a strong
likelihood that a counter-alignment torque acts on the inner
edge of a disk.  Again, we shall not attempt to consider the
complex and detailed evolution of a disk under these torque
components, but restrict our discussion to the question of
whether or not they could cause, in principle, time-varying
phenomena that are observable.

With a sign reversal of $\delta\Gamma_{a}$, equation (27) gives
a time constant $t_{a}$ for exponential growth of $\beta$,
\begin{eqnarray}
\frac{1}{t_{a}} \approx \frac{\mathcal{L}_{w}}
{8\pi\Omega\Sigma(GM)^{1/2}r_{i}^{5/2}}.
\end{eqnarray}
Evaluation for 4U 0142+61, with the parameters of Wang et al and
a local density $\Sigma = 300$ g cm$^{-2}$, gives
$t_{a} \approx 1.1 \times 10^{13}$ s, which is too long to be of
observable interest.  But the value of $\mathcal{L}_{w}$ is small
for this neutron star and the inner radius of the disk is large. 
Under the assumption that, for example, PSR 1828-11 has a disk
external to its light cylinder we find, given the ATNF parameters
for this pulsar, that growth times could be quite short,
$t_{a} \approx 7\times 10^{6}(r_{i}/R_{LC})^{5/2}$ s.   

\section{Conclusions}

We have found that the azimuthal component of the pulsar wind
momentum-density exerts an $r$-dependent torque on a thin dust
or debris disk external to the light cylinder.  Both the
Poynting and relativistic particle components of the wind are
responsible for ablation of the disk.  For disks that are
optically thick to relativistic protons, the power input, which
we assume determines the infra-red luminosity, is given by
equation (15).  This result neglects any albedo, principally,
in this case, reflection of ${\bf E}$ and ${\bf B}$ fields at
the disk surface which could be estimated only by a much more
complete treatment of this problem than we have attempted.
The rate of ablation is given by equation (16) in terms of the wind
luminosity $\mathcal{L}_{w}$ and the cascade parameters
$\mathcal{E}_{p}$
and $\mathcal{N}$ for which we have only qualitative estimates.
Under the very loose assumption that $\Omega_{0}\sin\beta$ is
a universal constant for all isolated neutron stars that are
formed with dust or debris fall-back disks, equation (18) shows
that $r_{i}^{3}\bar{\Sigma}$ should also be a universal constant.
From the parameters given for 4U 0142+61 by Wang et al, its value
is $1.3\times 10^{38}$ g cm.

It is obvious that our assumption of a common formation spin
angular frequency $\Omega_{0}$ for all neutron stars should not
be taken too seriously.  But if the  4U 0142+61 analysis of Wang
et al and our analysis of disk ablation are both accepted,
equation (18) makes it possible to infer what disk parameters
are possible in other isolated neutron stars and pulsars.
The conclusion we reach on this basis is that long-lived disks
of the kind considered here could be present in many pulsars
without exceeding published limits on infra-red luminosity.

It has not been possible to answer with any certainty the question
of whether or not the torque components calculated here could
be responsible for any observable time-varying phenomena.
PSR 1828-11 is an interesting case of a pulsar that appears
to have periodic variations in timing residuals and in pulse
shape consistent with small-amplitude Eulerian precession
of its spin axis relative to coordinates fixed in the star
(Stairs, Lyne \& Shemar 2000).
If the inferred $500$d period were equated with the growth time
$t_{a}$
estimated at the end of Section 3.3 for counter-alignment of
the inner edge of a disk, the required radius would be
$r_{i} = 2.1 R_{LC}$ but the mean disk density would have to
$\bar{\Sigma} \approx 2 \times 10^{9}$ g cm$^{-2}$. An almost
identical
value would be required in the radio pulsar, PSR B1642-03,
which also appears to have periodic timing residuals
(Shabanova, Lyne \& Urama 2001). This might be thought a high
density for a nominally thin disk but, assuming, for example, an
outer radius $r_{o} = 2r_{i}$, the mass is only
$M_{d} = 1.5\times 10^{-4} M_{\odot}$, well below observed upper
limits.  In principle, movement of the ionized inner edge may have
the potential to produce small changes in the observed
radio-frequency emission pulse shape
and in the pulsar spin-down rate.  But it is not obvious why the
changes should be periodic, as observed.

The more recent case of the isolated neutron star
RX J0720.4-3125 in which the inferred blackbody temperature
and certain other pulse characteristics have been interpreted as
varying sinusoidally is rather different (Haberl et al 2006; but
see also van Kerkwijk et al 2007 for an alternative analysis).  This
neutron star with a period of $8.39$ s and a high magnetic field
has some properties in common with 4U 0142+61.  The small
luminosity, $\mathcal{L}_{w} = 4.7\times 10^{30}$ erg s$^{-1}$
(Manchester et al 2005), is such that, even for $r_{i} = R_{LC}$,
the variation-time estimate obtained from equation (28) is
several orders of magnitude too long to be of interest.

\section*{Acknowledgments}

It is a pleasure to thank the referee, Professor M. A. Alpar,
for some very helpful comments on this paper.

\bsp

\label{lastpage}

\end{document}